# Dirac Cone in Two Dimensional Bilayer Graphene by Intercalation with V, Nb, and Ta Transition Metals

Srimanta Pakhira,[1,2,3] Kevin P. Lucht,[2,3] and Jose L. Mendoza-Cortes[1,2,3,*]

[1]*Condensed Matter Theory, National High Magnetic Field Laboratory,
Florida State University, Tallahassee FL, 32310, USA*
[2]*Scientific Computing Department, Materials Science and Engineering,
Florida State University, Tallahassee FL, 32310, USA*
[3]*Department of Chemical & Biomedical Engineering, FAMU-FSU Joint College of Engineering,
Florida State University, Tallahassee FL, 32310, USA.*

(Dated: September 16, 2017)

Bilayer graphene (BLG) is semiconductor whose band gap and properties can be tuned by various methods such as doping or applying gate voltage. Here, we show how to tune electronic properties of BLG by intercalation of transition metal (TM) atoms between two monolayer graphene (MLG) using a novel dispersion-corrected first-principle density functional theory approach. We intercalated V, Nb, and Ta atoms between two MLG. We found that the symmetry, the spin, and the concentration of TM atoms in BLG-intercalated materials are the important parameters to control and to obtain a Dirac Cone in their band structures. Our study reveals that the BLG intercalated with one Vanadium (V) atom, BLG-1V, has a Dirac Cone at the K-point. In all the cases, the present DFT calculations show that the $2p_z$ sub-shells of C atoms in graphene and the $3d_{yz}$ sub-shells of the TM atoms provide the electron density near the Fermi level $E_F$ which controls the material properties. Thus, we show that out-of-plane atoms can influence in-plane electronic densities in BLG, and enumerate the conditions necessary to control the Dirac point. This study presents a new strategy for controlling the material properties of BLG so that they exhibit various behaviors, including: metal, semi-metal, and semiconductor by varying the concentration and spin arrangement of the TM atoms in BLG while offering insight into the physical properties of 2D BLG-intercalated materials.

## I. INTRODUCTION

In modern science and technology, there has been a tremendous amount of theoretical and experimental interest in the low energy electronic properties of ultra-thin graphite films including graphene monolayer and graphene bilayer materials. [1–6] Graphene is a 2D honeycomb sheet of carbon, just one atom thick, with hybridized sp$^2$ bonded orbitals between carbon atoms. [7, 8] Graphene is an ideal material for making nanoelectronic and photonic devices because it is a very good electrical conductor as well as the thinnest 2D material known. Graphene has a unique, linear band structure around the Fermi level ($E_F$) forming a Dirac Cone at the K-points of its Brillouin zone. This has led to fascinating phenomena, exemplified by massless Dirac fermion physics.[9–12] This emergent behavior of Dirac fermions in condensed matter systems defines the unifying framework for a class of materials called Dirac materials.

Monolayer graphene (MLG) or simply graphene, has an electronic structure that can be controlled by an electrical field.[13] However, MLG has the well known zero-gap issue [14] which makes a high on-off ratio difficult; deeming it unsuitable for transistors, which are the foundation of all modern electronic devices. Therefore, opening the band gap of graphene still remains a challenging task to date. Bilayer graphene (BLG) can be used instead of MLG to overcome the zero-gap problem, with a gap opening simply by applying an electric field.[15, 16] BLG has an entirely different band structure, though it is equally interesting.[17, 18] Since the band gap of the BLG can be modulated from zero to a few eV by using different methods, such as doping and applying an external electric field. In addition, BLG holds the potential for electronics and nano-technological applications, particularly because of the possibility of controlling both carrier density and the energy band gap through doping or gating. [9, 10, 18–20] The most remarkable property of BLG is that the inversion symmetric AB-stacked BLG is a zero-band gap semiconductor in its pristine form, but a non-zero band gap can be induced by breaking the inversion symmetry of the two graphene monolayers to form AA-stacked BLG.[18–20] When two graphene monolayers are stacked (in both the cases AA- and AB-stacked), the monolayer features are lost and dispersion effects become quadratic and more effective. [19] Thus, BLG acts as a semiconductor and exhibits an induced electric field and broadly tunable band gap.[10, 21] By tuning the band gap, BLG can transform from a semiconductor into a metal or semi-metal, which means that a single millimeter-square sheet of BLG could potentially hold millions of differently tuned electronic devices that can be reconfigured. Recently, lasers have been used to get BLG to act as: a conductor, a Dirac material, or a semiconductor. This is an important step toward the invention of computer chips made of a 2D material.[22]

[*] Corresponding Author: Prof. J. L. Mendoza-Cortes, email: jmendozacortes@fsu.edu

Metals interacting with MLG and BLG have been investigated both experimentally as well as theoretically by first principles computational methods. Novoselov and co-workers experimentally investigated detailed surface interactions between MLG and three metals (Au, Fe, and Cr) using scanning transmission electron microscopy.[23] Intercalation occurs when metal atoms are inserted between two graphene monolayers, e.g. BLG. Accordingly, Takahashi and co-workers [24] experimentally fabricated Li-intercalated BLG on a SiC(0001) substrate and characterized the crystallographic and electronic properties by low-energy electron diffraction (LEED) and angle-resolved photoemission spectroscopy (ARPES). They predicted that Li-intercalation in BLG would open a way to develop a nano-scale ion battery. Wang et al.[25] studied the energetics of the intercalation of alkali metal atoms (Li, Na and K) in graphite by van der Waals density functionals. They found that the intercalation induces only a subtle increase of in-plane C-C bond lengths, with longer C-C bonds in the vicinity of the alkali metals. Additionally, intercalation was accomplished but without breaking the hexagonal symmetry. Very recently, Hasegawa and co-workers found direct evidence for superconductivity in Ca-intercalated BLG using *in situ* electrical transport measurements. [26] To date, only a few computational studies have examined the electronic properties of BLG intercalated with transition metals (TMs). For example, intercalation of BLG with Cr[27, 28] and Mn[29] are predicted for spintronic applications. A theoretical investigation of BLG intercalated with Cr was carried out using DFT, which found that BLG-Cr layer material is ferromagnetic.[27] Using first-principles calculations, Liao et al.[30] investigated various stable BLG structures intercalated with Sc and Ti. They found that the strain was important for the stability of high-coverage TM in BLG. The effect of TM on the electronic and magnetic properties of defective BLG was investigated using DFT.[31] In spite of these reports, the structures and electronic properties are not well studied yet for BLG intercalated with TM from the same group column in the periodic table, specifically: V, Nb, and Ta. In this work, we apply first-principles calculations to search for new BLG intercalated with V, Nb, and Ta atoms. Here, we show that the Dirac Cone in the band structure of BLG can be controlled by adding TM atoms between two graphene layers (where TM = V, Nb, and Ta) so that the electronic band gap between the valence and conduction bands can be tuned; thus resulting in the appearance of a Dirac Cone only in BLG-1V. We found that intercalation of TM atoms, and their spin arrangement, causes a surprisingly diverse array of electronic properties. Additionally, the band gap depends on the coupling between the two graphene layers and symmetry of the BLG system, i.e. AB stacking vs AA stacking. We have also performed a theoretical investigation of the electronic properties such as: band structure, density of states (DOSs), spin arrangements, and structural stability for MLG, BLG and BLG intercalated with one, two and three Vanadium atom; BLG-1V, BLG-2V and BLG-3V, as shown in Figure 1. Other intercalated materials presented here include BLG intercalated with Nb and Ta.

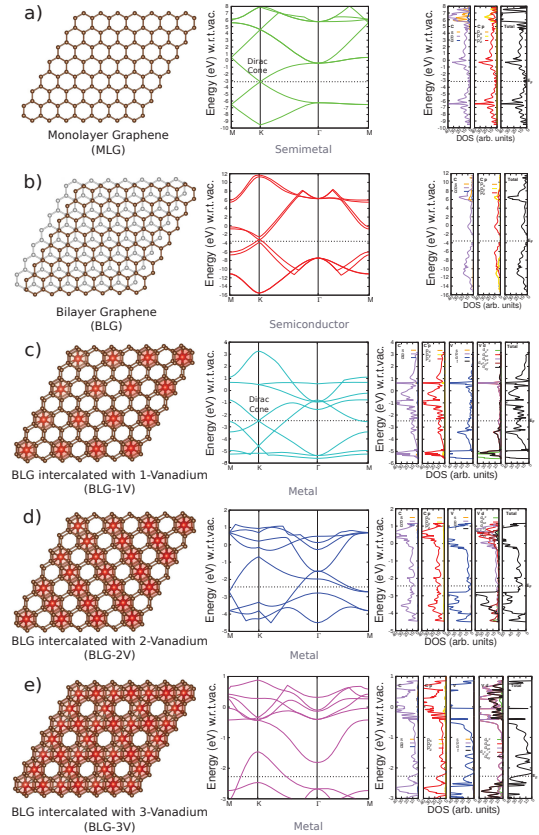

FIG. 1. The optimized structures, band structures and density of states (DOSs) are shown a) MLG, b) BLG, c) BLG-1V, d) BLG-2V, and e) BLG-3V. The individual components of the DOSs of the C and V atoms and total DOSs (depicted by "Total") are also presented in here. The energy for the band structures and DOSs is reported with respect to vacuum (i.e. w.r.t. vac.).

## II. COMPUTATIONAL DETAILS

The geometry of the 2D layer structures MLG, BLG and BLG-intercalated materials (BLG-1V, BLG-2V, BLG-3V, BLG-1Nb and BLG-1Ta) were optimized using dispersion-corrected unrestricted hybrid density functional theory, [32–37] i.e. UB3LYP-D2 which has been shown to give correct electronic properties for 2D materials.[35, 38–41] The semi-empirical Grimme-D2 dispersion corrections were added in the present calculations in order to incorporate van der Waals dispersion effects on the system.[35–37] For simplicity, this UB3LYP-D2 method is termed as DFT-D through out this work. The CRYSTAL14[42] suite code, which makes use of localized Gaussian basis sets, was used to perform all of the first-principles computations. During the optimization and single point energy calculations, 'SPIN' and 'SPIN-LOCK' keywords were used to specify the unrestricted wave functions and the total spin of the BLG-intercalated materials, respectively. The 'ATOMSPIN' keyword was also used to specify the individual spins of the TM atoms in the BLG-intercalated materials. The ferromagnetic (FM) and anti-ferromagnetic (AFM) spin alignment were studied by using the equivalent spin polarized wave function for Gaussian basis sets. The bilayer graphene intercalated 2D materials (BLG-1V, BLG-2V, BLG-3V and BLG-1Nb and BLG-1Ta) have been prepared by adding the V, Nb and Ta atoms in one unit cell of the crystal structures. In the present computation, triple-zeta valence with polarization (TZVP) quality Gaussian basis sets were used for both C and V, [43] and HAYWSC effective core potentials (ECP) were used for both Nb and Ta. [44] The threshold used for evaluating the convergence of the energy, forces, and electron density was $10^{-7}$ au for each parameter. The height of the cell was formally set to 500 Å, i.e. the vacuum region of approximately 500 Å was considered in the present calculations. The DFT exchange-correlation contribution was evaluated by numerical integration over the unit cell volume. Radial and angular points of the integration grid were generated through Gauss-Legendre radial quadrature and Lebedev two dimensional angular point distributions.

Integration inside of the first Brillouin zone were sampled on 15 × 15 × 1 k-mesh grids in all the materials for both the optimization and material properties (band structures and density of states) calculations with a resolution of around $2\pi \times (1/60)$ Å$^{-1}$ for all bilayer graphene intercalated materials, and $2\pi \times (1/40)$ Å$^{-1}$ for the MLG. In this work, we used a shrinking factor (the number of points along each reciprocal lattice vector at which the Fock matrix is diagonalized) of 30. We have plotted the bands along a high symmetric k-direction, M − K − Γ − M, in the first Brillouin zone. Electrostatic potential calculations have been included in the present computation, i.e. the energy for the band structures and DOSs is reported with respect to vacuum (i.e. w.r.t. vac.).

## III. RESULTS AND DISCUSSION

In this section, we present the results from DFT-D calculations (i.e. UB3LYP-D2) for the MLG, BLG, BLG-1V, BLG-2V, BLG-3V, BLG-1Nb, and BLG-1Ta layered materials. The present DFT calculation shows that the Dirac Cone exists at the K-point in the band structure of MLG, which is consistent with previous results.[9–12] The valence and conduction bands of MLG arise from the $2p_z$ sub-shells in C atoms, which form the Dirac Cone (see Figure 1a). Our present computation shows that AB-stacked BLG is thermodynamically more stable than the AA-stacked BLG. The valence and conduction bands overlap at the K-point for both the AA- and AB-stacked BLG. However, both bands do not align perfectly with the $E_F$, thus, the AA- and AB-stacked BLG have no Dirac Cone (Figure S1 in supplementary material and Figure 1b, respectively), which agrees with previous experimental results, [45, 46] i.e. the AA-/AB-stacked BLG materials are ordinary non-zero/zero band gap semiconductors. The bond distances, lattice constants ($a$ and $b$), and intercalation distance ($d$) are reported in Table I. The distance between the two monolayers in the pristine AB-stacked BLG material is about 3.390 Å which agrees with the previous experimental and theoretical results. [45] We have also calculated the binding energy ($\Delta E_f$) of all the systems studied here as shown in Table II. The binding energy, $\Delta E_f$, to form the AB-stacked BLG from two MLG, is about -0.86 eV, which is in agreement with the previous experimental and theoretical results.[47–50]

TABLE I. Bond distances and lattice constants for MLG, BLG and BLG-TM materials (where TM is transition metal). The $d$ is defined as the interlayer distance between the monolayers.

| Component | C-C (Å) | C-V (Å) | $a$ (Å) | $b$ (Å) | $d$ (Å) |
|---|---|---|---|---|---|
| MLG | 1.416 | N/A | 2.451 | 2.451 | N/A |
| AA-stacked BLG | 1.414 | N/A | 2.449 | 2.449 | 3.201 |
| AB-stacked BLG | 1.421 | N/A | 2.449 | 2.449 | 3.390 |
| BLG-1V | 1.439 | 2.243 | 4.942 | 4.942 | 3.441 |
| BLG-2V | 1.442 | 2.335 | 8.575 | 5.022 | 3.645 |
| BLG-3V | 1.464 | 2.413 | 5.032 | 5.032 | 3.658 |
| BLG-1Nb | 1.439 | 2.313 | 4.950 | 4.950 | 3.600 |
| BLG-1Ta | 1.445 | 2.284 | 4.952 | 4.952 | 3.522 |

TABLE II. Relevant properties of the MLG, BLG, and BLG-TM; the binding energy, $\Delta E_f$, is with respect to their individual components: MLG and TM atoms.

| Component | State | $\Delta E_f$ (eV) | Dirac Cone |
|---|---|---|---|
| MLG | Semimetal | N/A | Yes |
| AA-stacked BLG | Semiconductor | -0.78 | No |
| AB-stacked BLG | Semiconductor | -0.86 | No |
| BLG-1V | Metal | -5.97 | Yes |
| BLG-2V | Metal | -5.67 | No |
| BLG-3V | Metal | -5.36 | No |
| BLG-1Nb | Metal | -11.18 | No |
| BLG-1Ta | Metal | -16.35 | No |

The optimized structure between Vanadium and graphene follows an AA-stacking arrangement with the Vanadium placed at the center of the honeycomb, to form BLG-1V. In this conformation, the Vanadium $d$-orbitals are well situated to favorably interact with the $p_z$ orbital orthogonal to the graphene layer. The BLG-1V structure is highly favorable by -5.97 eV relative to BLG and one free V atom. Additional Vanadium atoms are also favorable for further intercalation by -5.67 eV between single and double metal addition to form BLG-2V, and by -5.36 eV between the double and triple metal addition to form BLG-3V as shown in Table II. The present DFT calculations show that the intercalation distances gradually increased when more V atoms are added from BLG-1V to BLG-3V, but not equally as shown in Table I. Our computation also shows the C-C bond distance has been increased by a cumulative average of 0.014 Å in BLG-1V, BLG-2V and BLG-3V due to the presence of V atoms between the graphene layers compared to BLG. We also calculated the stability of BLG intercalated with four V atoms (BLG-4V), but the frequency calculations show that the structure is unstable thermodynamically as it has many imaginary frequencies, and hence this result is excluded from this article.

The addition of Vanadium atoms was found to substantially alter the electronic properties of graphene bilayer. The addition of a single Vanadium atom intercalated in AA-stacked BLG (i.e. BLG-1V) yields a Dirac Cone along the K-direction as shown in Figure 1c. The individual components of the $p$-orbitals electron density for C atoms and the $d$-orbitals electron density for V atoms have been explicitly shown along with the total DOSs in Figure 1c. The DOSs calculations for BLG-1V show electron density around the $E_F$ indicating metallic behavior, and is coming from the $p_z$ sub-shells of C atoms in graphene and the $d_{yz}$ sub-shell of the V atom. In other words, the $p_z$ sub-shell of carbon, and $d_{yz}$ sub-shells of vanadium is mostly responsible for the emergence of the Dirac Cone in BLG-1V.

In MLG, the Dirac cone comes from the $p_z$ sub-shells of $p$-orbitals of the C atoms alone. However, for BLG-1V, the nature of the Dirac cone might be different because it might come from the $p$-orbitals alone in C, or $d$-orbitals alone of the V atom, or $p-d$ hybridized orbitals of C and V atoms. To resolve this, we used the coordinates of the optimized geometry of BLG-1V, and then (i)- removed the BLG, and in another calculations (ii)- removed the V atoms. We then calculated the electronic properties of these structures (Figure S2 in supplementary material). The band structure shows that the Dirac cone disappears when only the V atoms are used and also when only BLG is used. In the BLG case the Dirac point was still present but it shifted above $E_F$, just like in the ideal AA-stacked BLG. This suggests that $p-d$ hybridized orbitals are the source of the Dirac cone in BLG-1V. As an excellent conductor and Dirac material, BLG-1V can be used in various modern electronic devices.

BLG-2V and BLG-3V can be formed with the addition of one V atom and two V atoms to the BLG-1V material, respectively. Different spin states of BLG-2V and BLG-3V have been considered but only the most stable configurations are discussed here, while the other less stable spin states are discussed in the SI; see Table S1. First, we will discuss the Ferromagnetic (FM) state for both materials, which results in the Dirac Cone disappearing, and the bands shifting, which shows a metallic behavior when the spins of Vanadium atoms align. Interestingly, in general for all these materials in the FM state, the $3d$ orbitals of V receive more electron donation from the graphene $2p_z$ sub-shells when the total spin is increased. The band structure reveals that along the M to K directions, the valence band crosses the Fermi level in both BLG-2V and BLG-3V, resulting in large electron distribution around the Fermi level as shown in Figure 1d and Figure 1e, respectively. Thus, both BLG-2V and BLG-3V show metallic behavior as depicted in their total DOSs. In both materials, the total DOSs show that electron density around the Fermi level is due to the $p_z$ sub-shell electrons from C atoms in graphene and the $d_{yz}$ sub-shell electrons from V atoms as depicted in the sub-shells DOSs calculations (see Figure 1d - 1e).

The other relevant arrangement for both BLG-2V and BLG-3V is the anti-ferromagnetic (AFM) state, which is now briefly discussed. By considering the AFM arrangement of spins, we can drastically modify the properties of BLG-2V. The spin conformations of all the BLG-intercalated materials are reported in Table S1. Figure 2 shows the band structure and DOSs of the AFM state for BLG-2V. To further analyze the wave function, we calculated a Mulliken spin population analysis. In the case of BLG-2V, the AFM conformation yielded higher spin of individual V atoms, 2.22 and -2.22 compared to the FM conformation, which had a value of 1.86 for each Vanadium. Relative to its FM counterpart, the AFM BLG-2V structure is more stable by $\Delta E_f = 0.410$ eV. Examining the electronic properties, we noticed considerable differences in BLG-2V between its AFM and FM states. Once in the AFM state, BLG-2V has a degenerate pair of band structures (for alpha and beta electrons) with a band opening between the valence and conduction bands. The band gaps are 0.101 eV for the indirect transition and 0.681 eV for the direct transition. The most important finding is that by altering the spin to generate the AFM state in BLG-2V, this material becomes a semiconductor, which may be useful for modern electronic devices. Thus, this calculation reveals that a band gap can be introduced in BLG-2V by altering the spin alignment of the intercalated V atoms. The AFM state for BLG-3V was also found, but it was not as interesting as the AFM state of BLG-2V (Figure S3).

We further extended our studies on the equilibrium structures and electronic properties of Niobium (Nb) and Tantalum (Ta) intercalated BLG materials (BLG-1Nb and BLG-1Ta) to investigate if the Dirac Cone exists in these two materials; see Figure 3. The structural parameters of both materials are reported in Table 1. The DFT calculations found that the intercalation distance $d$ was increased in both materials compared to BLG-1V by at least 0.1 Å, which is consistent with Nb and Ta being larger than V. However, the interaction between the Nb or Ta and graphene layers is stronger than in BLG-1V, as suggested by the $\Delta E_f$ shown in Table II. The electronic properties of both materials are shown in Figure 3. The band structure calculations showed that they have no Dirac Cone; instead, the Dirac point moved above the $E_F$. The total DOSs calculation found that the electron densities were large around the $E_F$, which arises from the $p$-subshells of C atoms and $d$-subshell of Nb and Ta atoms as depicted in Figure 3. These calculations also showed that they have metallic behavior. In the present study, we considered only one Nb and Ta atom intercalated in BLG, because we showed that if the concentration of TM is increased, the Dirac Cone disappears (see BLG-2V and BLG-3V band structures in Figure 1d-e). Therefore, we excluded the investigation on BLG-2Nb, BLG-3Nb, BLG-2Ta and BLG-3Ta from this article.

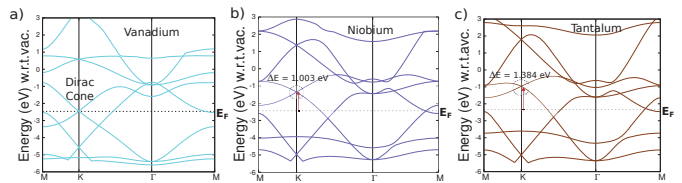

FIG. 4. Comparison of band structures of a) BLG-1V, b) BLG-1Nb, and b) BLG-1Ta showing how the Dirac point remains, but is shifted up with respect to the $E_F$.

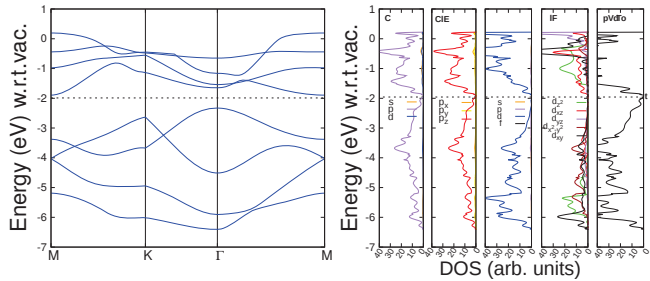

FIG. 2. Band structure and DOSs of the alpha electron in AFM state for BLG-2V are presented. The beta electrons exhibit identical information. The individual and total components of DOSs of the C and V atoms are also shown. Compared to the FM state, in the AFM state a band gap appears.

We have investigated why the Dirac Cone does not exist in BLG-1Nb and BLG-1Ta but the Dirac point still does. The BLG-1V, BLG-1Nb, and BLG-1Ta materials have the $p6/mmm$ layer symmetry, which is the same as monolayer graphene. The Mulliken spin population analysis found that the total spin of BLG-1V is 2.082, with vanadium having a spin of 2.245. Whereas, the total spins of BLG-1Nb and BLG-1Ta are 1.399 and 1.294, with niobium and tantalum having spins of 1.245 and 1.068, respectively. These calculations suggest that the number of unpaired electrons decreased when we intercalated heavier atoms such as Nb and Ta in bilayer graphene. As a consequence, the Dirac cone moved away from the Fermi level, as shown in Figure 4. The present DFT calculations found the band crossing points in the band structures of BLG-1Nb and BLG-1Ta were shifted by 1.003 eV and 1.384 eV, respectively as depicted in Figure 4. Thus, the individual unpaired electrons of V, Nb, and Ta play an important role the existence of a Dirac Cone in the band structures of the BLG intercalated materials. The BLG-2V material lost the $p6/mmm$ layer symmetry; and therefore, the Dirac point can not be present. Additionally, since the concentration of V atoms is increased, the metal features dominate (Figure 1d). Apart from the spin and symmetry, the concentration of the metal atoms in the BLG system is another important factor. BLG-3V has the same symmetry ($p6/mmm$) but the concentration of V atoms increased; thus, the symmetry allowed it to have graphene features. These features do not show up around the $E_F$, because mostly the TM properties dominate. In other words, it loses the Dirac Cone because of the dominance of the TM concentration, as displayed in Figure 1e.

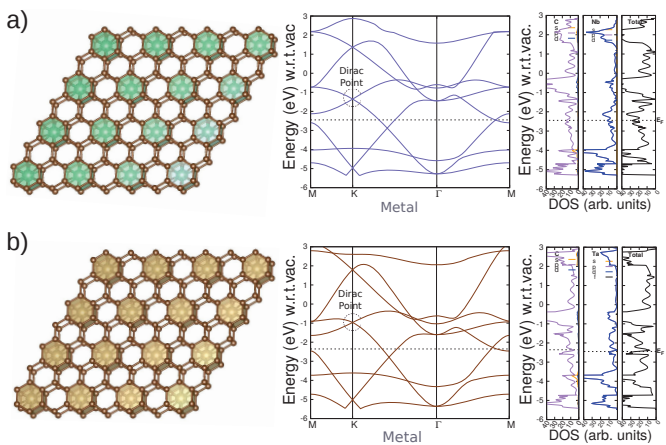

FIG. 3. The optimized structures, band structures, and density of states (DOSs) are shown for a) BLG-1Nb, and b) BLG-1Ta. The individual components of the DOSs of the C, Nb, and Ta atoms and total DOSs (depicted by "Total") are also presented in here.

## IV. CONCLUSIONS

We have studied the electronic structure and properties of the 2D-materials: MLG, BLG, and BLG intercalated with transitions metals (TM), where the TM = V, Nb and Ta. The individual components of the $p$-subshell of C atoms and the $d$-subshell of the V, Nb, Ta atoms are reported along with the total DOSs. These materials have favorable binding energy, thus, they might be feasible to synthesize experimentally. Among all the systems, MLG and BLG-1V materials have a Dirac Cone at the K-point. Once the concentration of Vanadium is increased the Dirac Cone disappears. We have also found that the $2p_z$ sub-shell of C atoms and $3d_{yz}$ sub-shell of the V atoms in the BLG-1V, BLG-2V, and BLG-3V materials are the main components around the $E_F$. This plays the main role in the presence the Dirac Cone and the conducting properties. Thus, we have shown that the control of the Dirac Cone on the intercalated BLG is a delicate balance between the amount of unpaired electrons, concentration of the TMs, and symmetry. This means that the layer symmetry ($p6/mmm$) is necessary for the Dirac point to appear in graphene-like materials (BLG-1V, BLG-1Nb and BLG-1Ta), however the $E_F$ is controlled by the number of unpaired electrons. In other words, if the concentration of TMs is too high, even having the same symmetry ($p6/mmm$) will not be sufficient to keep the Dirac point at the $E_F$, because the band features will be dominated by the TM. This is clearly shown in how the material properties have been changed in BLG-1V and BLG-3V due to the presence of Vanadium atoms with the same symmetry. The nature of the Dirac cone in MLG and BLG-1V is different with the former having $p$-orbitals character while the second involves $p$- and $d$- orbitals. These materials show how to control a Dirac point and tune it to become a Dirac cone. Thus, this work presents a promising approach to create more Dirac 2D-materials.

**Supplementary Materials** Supporting information and supplementary data related to this article can be found at "Supplementary Materials" file. See supplementary material for detailed (a) optimized geometry, band structure, and density of states of AA-stacked bilayer graphene; (b) Nature of the Dirac Cone in BLG-1V; (c) Effect of spin alignment on BLG-3V material; (d) Optimized Structures (.cif format).

**Acknowledgements** S.P. and J.L.M-C. were supported by Florida State University (FSU) and its Energy and Materials Initiative. S.P. is grateful to Dr. Y. Pramudya, Stephanie Marxsen and Oluwagbenga Oare Iyiola from FSU for helpful discussions and guidance with computational resources. The authors thank the High Performance Computer cluster at the Research Computing Center in FSU, for providing computational resources and support.

# Supplementary Material

# Dirac Cone in two dimensional bilayer graphene by intercalation with V, Nb, and Ta transition metals

Srimanta Pakhira[1,2,3], Kevin P. Lucht[2,3], Jose L. Mendoza-Cortes[1,2,*]

[1] Condensed Matter Theory, National High Magnetic Field Laboratory (NHMFL), Florida State University, Tallahassee, Florida, 32310, USA.

[2] Scientific Computing Department, Materials Science and Engineering, High Performance Materials Institute (HPMI), Florida State University, Tallahassee, Florida, 32310, USA.

[3] Department of Chemical & Biomedical Engineering, Florida A&M University - Florida State University, Joint College of Engineering, Tallahassee, Florida, 32310, USA.

E-mail: mendoza@eng.famu.fsu.edu

# Contents

# List of Figures

# 1 Optimized geometry, band structure and density of states of AA-stacked bilayer graphene

The optimized structure, band structures, and density of states (DOSs) of AA-stacked bilayer graphene (BLG) are shown in Figure S1. The band and DOSs calculations reveal that AA-stacked BLG is a non-zero band gap semiconductor, and it has an indirect band gap around 0.25 eV as depicted in the band structure. Notice that the band structure does not show the gap, but the band gap is reported based on the DOS calculations. The individual components of the p-orbitals (i.e. $p_x$, $p_y$ and $p_z$ sub-shells) are calculated along with the total DOSs. We found the $p_z$ sub-shell of the p-orbitals accounts for the largest electron contribution in the total DOSs.

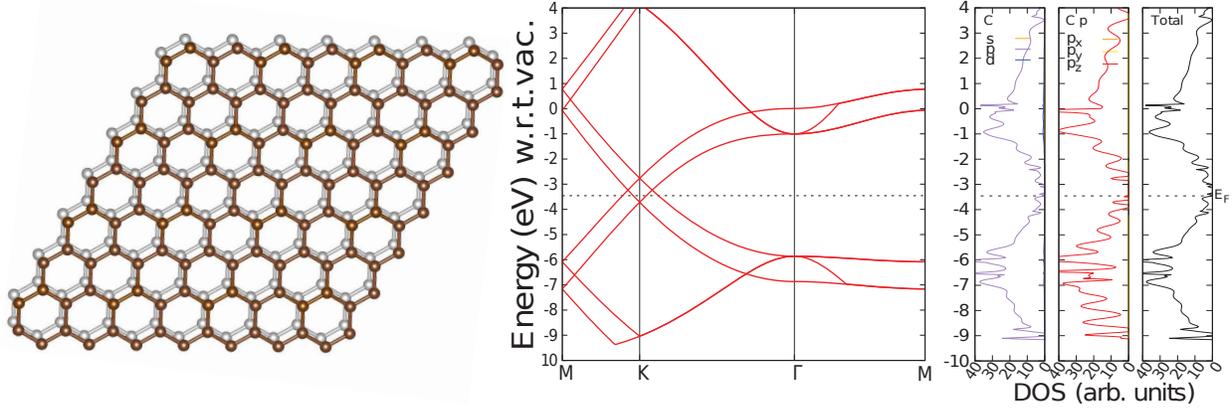

Figure S1: The optimized structure, band structure and density of states (DOSs) of AA-stacked BLG are shown. The individual components of DOSs of the C atom, and total DOSs (depicted by "Total" in the third column) are also presented.

The mathematical expressions that were used to calculate the binding energy of the pristine BLG and TM-intercalated BLG, $\Delta E_f$, are given below:

$$\Delta E_f = E_{BLG} - E_{MLG} \text{ ....... (1)}$$
$$\Delta E_f = E_{Intercalated-BLG} - E_{BLG} - nE_{TM} \text{ ....... (2)}$$

Where $E_{BLG}$ is the energy of the pristine BLG, $E_{MLG}$ is the energy of the MLG, $E_{Intercalated-BLG}$ is the energy of the TM-intercalated BLG, $E_{TM}$ is the energy of TM atom(s) and $n$ is the number of TM atoms intercalated in BLG.

## 2 Nature of the Dirac Cone in BLG-1V

In MLG, the Dirac Cone comes from the p-obitals of C atoms alone, but for BLG-1V, the nature of the Dirac Cone might be different because it might come from the $p$-orbitals alone, or $d$-orbitals alone or $p-d$ hybridized orbitals. To resolve this, we use the optimized geometry of BLG-1V, and then 1.- removed the BLG layer and 2.- removes the V atoms. The results are shown in Figure S2. Notice how the Dirac Cone disappears when only the V atoms are used Figure S2b. The same occurs when only the BLG is used, Figure S2c; however, in this case, the Dirac Cone is present but has shifted above $E_F$. This results suggests that $p-d$ hybridized orbitals are the source for the Dirac cone in BLG-1V.

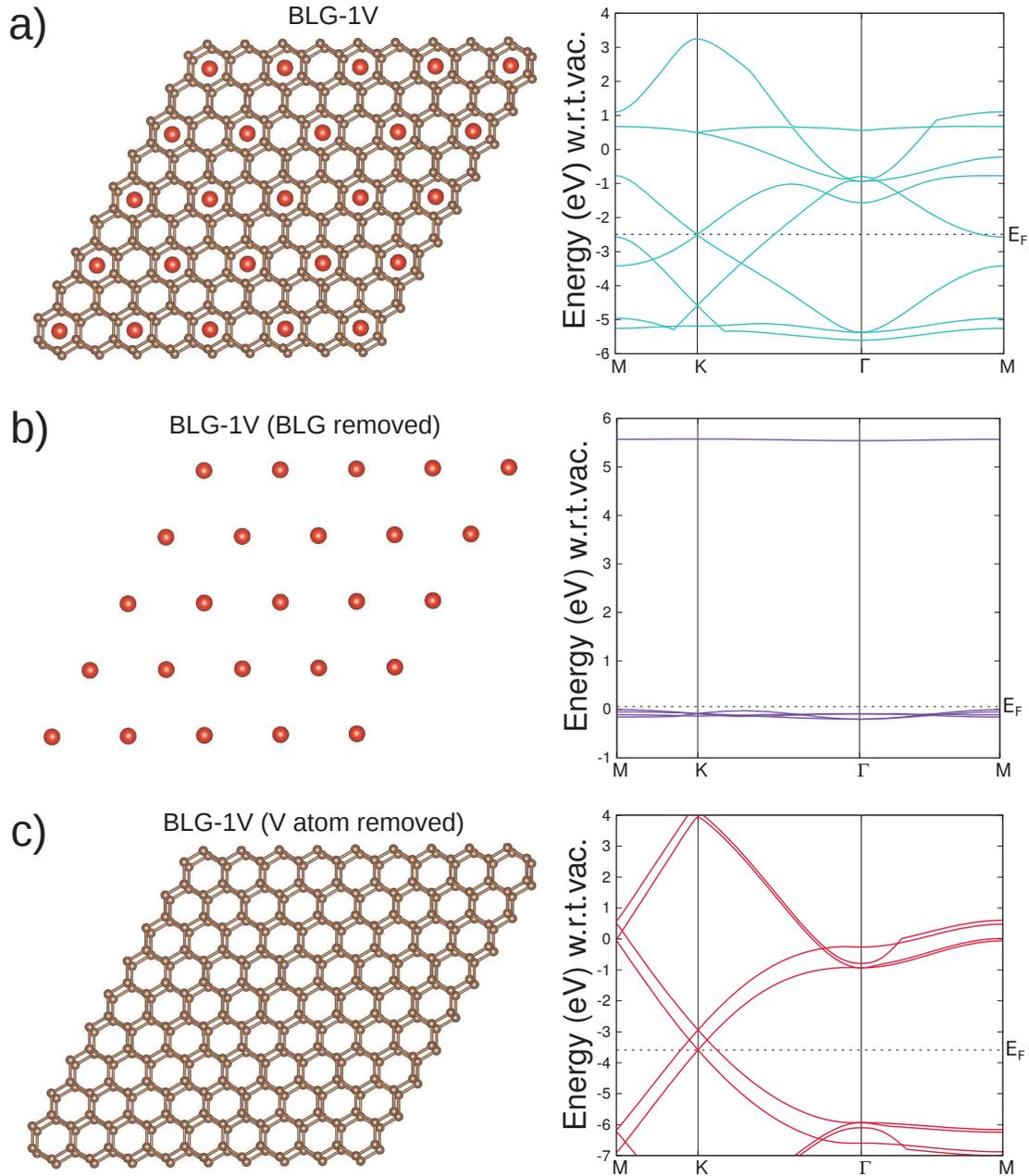

Figure S2: Electronic properties (left) and geometry (right) for a) BLG-1V, b) BLG-1V with the BLG removed and c) BLG-1V with the V atom removed

## 3 Effect of spin alignment on BLG-3V

An alternative method of controlling the electronic properties of BLG-intercalated materials is by altering the spin configuration. We carried out Mulliken spin density analysis to study the spin configuration at the UB3LYP-D2 level. All previous materials in this work have Vanadium as reference in the high-spin state, and BLG-2V and BLG-3V in a ferromagnetic (FM) spin arrangement. We can drastically modify the properties by considering an anti-ferromagnetic (AFM) arrangement of spins. The AFM arrangement of BLG-2V was reported in the main manuscript. Here we showed the spin conformations of the BLG-2V and BLG-3V materials, and the Mulliken spin densities of the V atoms in the BLG-intercalated materials are reported in Table S1.

Table S1: Different Mulliken spin population of the BLG-intercalated materials.

| Materials | Average Spin of V | Total Spin |
| --- | --- | --- |
| BLG-1V | 2.245 | 2.082 |
| FM BLG-2V | 1.856 | 3.740 |
| FM BLG-3V | 1.489 | 4.375 |
| AFM BLG-2V | 0.000 (2.215, -2.215) | 0.000 |
| AFM BLG-3V | 0.157 | 0.405 |
| BLG-1Nb | 1.245 | 1.399 |
| BLG-1Ta | 1.068 | 1.294 |

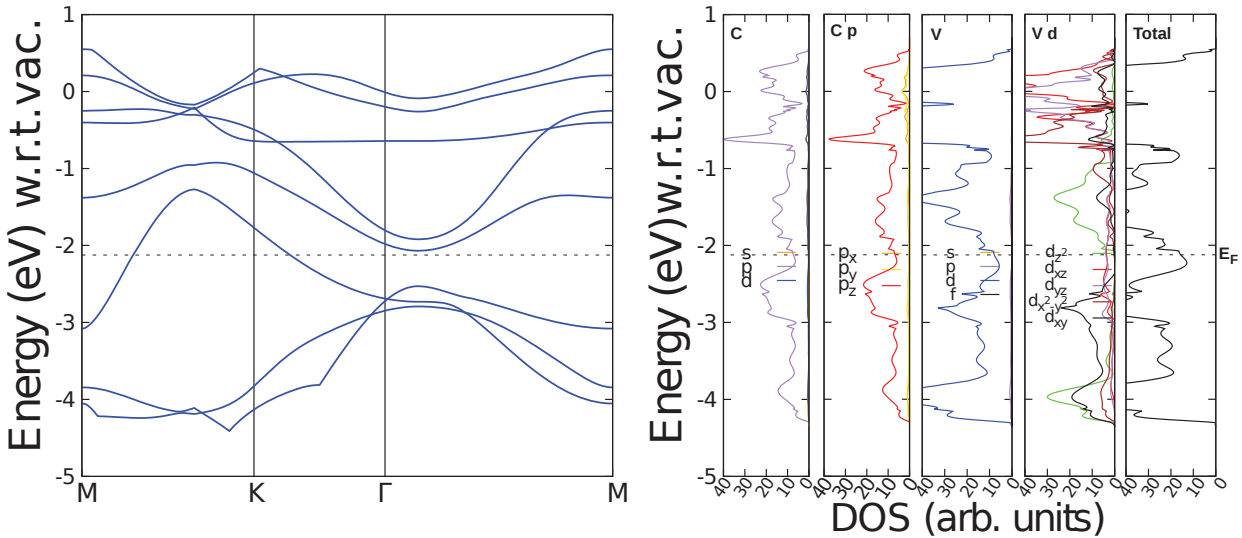

Figure S3: The band structure and DOSs of the alpha electron of AFM BLG-3V are presented. The individual components of the DOSs contributions from the C and V atoms, and total DOSs are also presented.

For the case of BLG-3V, there also exists one AFM conformation given the symmetry of the unit cell consisting of two Vanadium atoms with electrons in the alpha state and one in the beta state, or vice versa. We found that the AFM BLG-3V structure chooses to allocate the spin as -1.726 for the beta state Vanadium, and the two alpha state Vanadiums with spins of 1.098. The average spin of AFM BLG-3V is 0.405. The AFM structure of BLG-3V is slightly less favorable than to the FM structure by $\Delta G_f = 0.193$ eV. As for the electronic properties, we found that for alpha and beta electrons that it is consistent with being conductive based on their DOSs and band overlap around the Fermi Energy (see Figure S3 and Figure S4). The present calculations showed that both the FM and AFM states of BLG-3V are conducting and thus they show ordinary metallic behavior. This study of spin behavior shows that not only the electronic properties of BLG-intercalated materials dependent on the concentration of Vanadium atoms intercalated in BLG, but also on the arrangement of the spin, which adds an additional variable for modifying the behavior of BLG-intercalated materials.

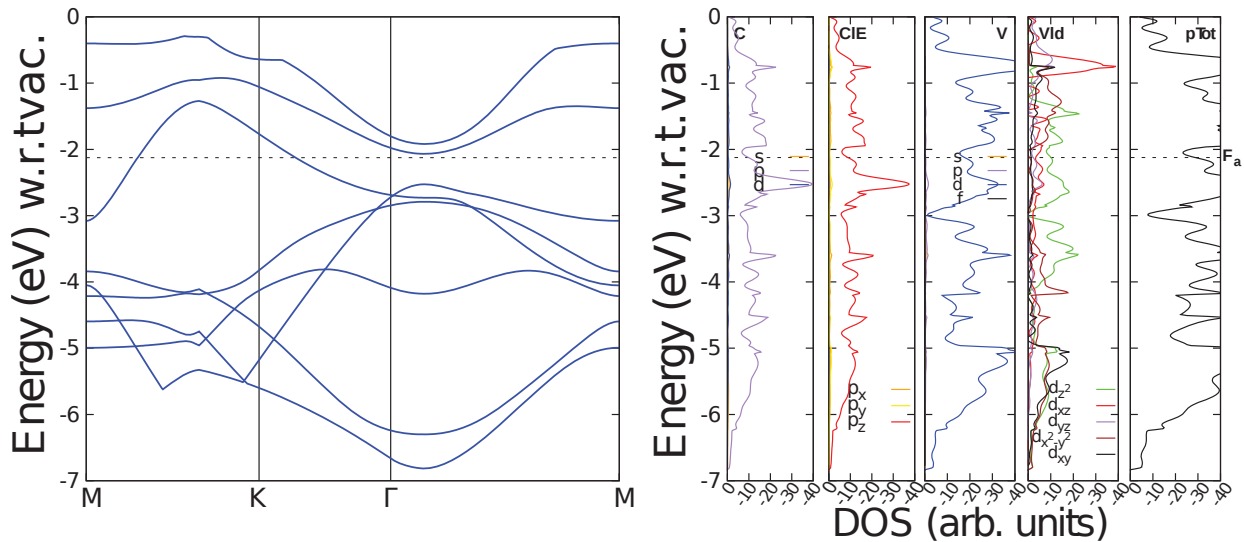

Figure S4: The band structure and DOSs of the beta electron of AFM BLG-3V are presented. The individual components of the DOSs contributions from the C and V atoms, and total DOSs are also presented here. This is consistent with conductive behavior for the alpha and beta electrons.

Thus, we show that an alternative method of controlling the electronic properties of BLG-intercalated materials is altering the spin conformation. The AFM arrangement of spins in the BLG-2V material changes it from a metal to a semi-conductor with a small band gap. For BLG-3V, the AFM state changes the electron density around the Fermi level compared FM states. We have observed that the $3d$ orbitals of V receive more electron donation from the graphene $2p_z$ sub-shell when the total spin is increased. We have found the $2p_z$ sub-shell of $p$-orbital of C atoms and $3d_{yz}$ sub-shell of d-orbital of V atoms account for the largest electron contribution in total DOSs.

# 4 Optimized Structures (.cif format)

The optimized structures are provided below in .cif format.

## 4.1 MLG: Monolayer Graphene

```
data_MLG
_symmetry_space_group_name_H-M    'P6/MMM'
_symmetry_Int_Tables_number       191
_symmetry_cell_setting            hexagonal
loop_
_symmetry_equiv_pos_as_xyz
  x,y,z
  -y,x-y,z
  -x+y,-x,z
  -x,-y,z
  y,-x+y,z
  x-y,x,z
```

```
  y,x,-z
  x-y,-y,-z
  -x,-x+y,-z
  -y,-x,-z
  -x+y,y,-z
  x,x-y,-z
  -x,-y,-z
  y,-x+y,-z
  x-y,x,-z
  x,y,-z
  -y,x-y,-z
  -x+y,-x,-z
  -y,-x,z
  -x+y,y,z
  x,x-y,z
  y,x,z
  x-y,-y,z
  -x,-x+y,z
_cell_length_a                 2.4515
_cell_length_b                 2.4515
_cell_length_c                 500.0000
_cell_angle_alpha              90.0000
_cell_angle_beta               90.0000
_cell_angle_gamma              120.0000
loop_
_atom_site_label
_atom_site_type_symbol
_atom_site_fract_x
_atom_site_fract_y
_atom_site_fract_z
_atom_site_U_iso_or_equiv
_atom_site_adp_type
_atom_site_occupancy
C001   C    0.33333   -0.33333   0.00000    0.00000   Uiso    1.00
```

## 4.2  AA-BLG: AA-stacked Bilayer Graphene

```
data_AA-Stacked-BLG
_symmetry_space_group_name_H-M    'P6/MMM'
_symmetry_Int_Tables_number       191
_symmetry_cell_setting            hexagonal
loop_
_symmetry_equiv_pos_as_xyz
  x,y,z
  -y,x-y,z
  -x+y,-x,z
  -x,-y,z
  y,-x+y,z
  x-y,x,z
  y,x,-z
  x-y,-y,-z
  -x,-x+y,-z
  -y,-x,-z
  -x+y,y,-z
  x,x-y,-z
```

```
  -x,-y,-z
  y,-x+y,-z
  x-y,x,-z
  x,y,-z
  -y,x-y,-z
  -x+y,-x,-z
  -y,-x,z
  -x+y,y,z
  x,x-y,z
  y,x,z
  x-y,-y,z
  -x,-x+y,z
_cell_length_a                 2.4490
_cell_length_b                 2.4490
_cell_length_c                 500.0000
_cell_angle_alpha              90.0000
_cell_angle_beta               90.0000
_cell_angle_gamma              120.0000
loop_
_atom_site_label
_atom_site_type_symbol
_atom_site_fract_x
_atom_site_fract_y
_atom_site_fract_z
_atom_site_U_iso_or_equiv
_atom_site_adp_type
_atom_site_occupancy
C001   C    -0.66667  -0.33333   0.00320   0.00000   Uiso   1.00
```

## 4.3 AB-BLG: AB-stacked Bilayer Graphene

```
data_AB-Stacked-BLG
_symmetry_space_group_name_H-M    'P-3M1'
_symmetry_Int_Tables_number       164
_symmetry_cell_setting            trigonal
loop_
_symmetry_equiv_pos_as_xyz
  x,y,z
  -y,x-y,z
  -x+y,-x,z
  y,x,-z
  x-y,-y,-z
  -x,-x+y,-z
  -x,-y,-z
  y,-x+y,-z
  x-y,x,-z
  -y,-x,z
  -x+y,y,z
  x,x-y,z
_cell_length_a                 2.4491
_cell_length_b                 2.4491
_cell_length_c                 500.0000
_cell_angle_alpha              90.0000
_cell_angle_beta               90.0000
_cell_angle_gamma              120.0000
```

```
loop_
_atom_site_label
_atom_site_type_symbol
_atom_site_fract_x
_atom_site_fract_y
_atom_site_fract_z
_atom_site_U_iso_or_equiv
_atom_site_adp_type
_atom_site_occupancy
C001  C    0.00000   0.00000   0.00304   0.00000  Uiso   1.00
C003  C   -0.33333   0.33333   0.00304   0.00000  Uiso   1.00
```

## 4.4  BLG-1V: BLG-intercalated with 1 V

```
data_BLG-1V
_symmetry_space_group_name_H-M    'P6/MMM'
_symmetry_Int_Tables_number       191
_symmetry_cell_setting            hexagonal
loop_
_symmetry_equiv_pos_as_xyz
  x,y,z
  -y,x-y,z
  -x+y,-x,z
  -x,-y,z
  y,-x+y,z
  x-y,x,z
  y,x,-z
  x-y,-y,-z
  -x,-x+y,-z
  -y,-x,-z
  -x+y,y,-z
  x,x-y,-z
  -x,-y,-z
  y,-x+y,-z
  x-y,x,-z
  x,y,-z
  -y,x-y,-z
  -x+y,-x,-z
  -y,-x,z
  -x+y,y,z
  x,x-y,z
  y,x,z
  x-y,-y,z
  -x,-x+y,z
_cell_length_a                    4.9423
_cell_length_b                    4.9423
_cell_length_c                    500.0000
_cell_angle_alpha                 90.0000
_cell_angle_beta                  90.0000
_cell_angle_gamma                 120.0000
loop_
_atom_site_label
_atom_site_type_symbol
_atom_site_fract_x
_atom_site_fract_y
```

```
_atom_site_fract_z
_atom_site_U_iso_or_equiv
_atom_site_adp_type
_atom_site_occupancy
C001  C   -0.33601  -0.16801   0.00344   0.00000  Uiso  1.00
C007  C    0.33333  -0.33333   0.00339   0.00000  Uiso  1.00
V009  V    0.00000   0.00000   0.00000   0.00000  Uiso  1.00
```

## 4.5 BLG-2V: BLG-intercalated with 2 V

```
data_BLG-2V
_symmetry_space_group_name_H-M     'CMM2'
_symmetry_Int_Tables_number        35
_symmetry_cell_setting             orthorhombic
loop_
_symmetry_equiv_pos_as_xyz
  x,y,z
  -x,-y,z
  x,-y,z
  -x,y,z
  x+1/2,y+1/2,z
  -x+1/2,-y+1/2,z
  x+1/2,-y+1/2,z
  -x+1/2,y+1/2,z
_cell_length_a                     5.0217
_cell_length_b                     8.5748
_cell_length_c                     500.0000
_cell_angle_alpha                  90.0000
_cell_angle_beta                   90.0000
_cell_angle_gamma                  90.0000
loop_
_atom_site_label
_atom_site_type_symbol
_atom_site_fract_x
_atom_site_fract_y
_atom_site_fract_z
_atom_site_U_iso_or_equiv
_atom_site_adp_type
_atom_site_occupancy
C001  C    0.24997   0.08385  -0.00854   0.00000  Uiso  1.00
C010  C    0.24996   0.08385  -0.00125   0.00000  Uiso  1.00
C005  C    0.00000  -0.16606  -0.00840   0.00000  Uiso  1.00
C007  C    0.00000  -0.33405  -0.00840   0.00000  Uiso  1.00
C014  C    0.00000  -0.16608  -0.00139   0.00000  Uiso  1.00
C016  C    0.00000  -0.33403  -0.00139   0.00000  Uiso  1.00
V009  V    0.00000   0.00000  -0.00490   0.00000  Uiso  1.00
V018  V    0.50000   0.00000  -0.00490   0.00000  Uiso  1.00
```

## 4.6 BLG-3V: BLG-intercalated with 3 V

```
data_BLG-3V
_symmetry_space_group_name_H-M     'P6/MMM'
_symmetry_Int_Tables_number        191
_symmetry_cell_setting             hexagonal
loop_
_symmetry_equiv_pos_as_xyz
```

```
  x,y,z
  -y,x-y,z
  -x+y,-x,z
  -x,-y,z
  y,-x+y,z
  x-y,x,z
  y,x,-z
  x-y,-y,-z
  -x,-x+y,-z
  -y,-x,-z
  -x+y,y,-z
  x,x-y,-z
  -x,-y,-z
  y,-x+y,-z
  x-y,x,-z
  x,y,-z
  -y,x-y,-z
  -x+y,-x,-z
  -y,-x,z
  -x+y,y,z
  x,x-y,z
  y,x,z
  x-y,-y,z
  -x,-x+y,z
_cell_length_a                    5.0319
_cell_length_b                    5.0319
_cell_length_c                    20.0000
_cell_angle_alpha                 90.0000
_cell_angle_beta                  90.0000
_cell_angle_gamma                 120.0000
loop_
_atom_site_label
_atom_site_type_symbol
_atom_site_fract_x
_atom_site_fract_y
_atom_site_fract_z
_atom_site_U_iso_or_equiv
_atom_site_adp_type
_atom_site_occupancy
C001   C    0.33575   0.16787   0.08988   0.00000   Uiso   1.00
C013   C   -0.33333   0.33333   0.09613   0.00000   Uiso   1.00
V017   V    0.50000   0.00000   0.00000   0.00000   Uiso   1.00
```

## 4.7   BLG-1Nb: BLG-intercalated with 1 Nb

```
data_BLG-1Nb
_symmetry_space_group_name_H-M    'P6/MMM'
_symmetry_Int_Tables_number       191
_symmetry_cell_setting            hexagonal
loop_
_symmetry_equiv_pos_as_xyz
  x,y,z
  -y,x-y,z
  -x+y,-x,z
  -x,-y,z
```

```
  y,-x+y,z
  x-y,x,z
  y,x,-z
  x-y,-y,-z
  -x,-x+y,-z
  -y,-x,-z
  -x+y,y,-z
  x,x-y,-z
  -x,-y,-z
  y,-x+y,-z
  x-y,x,-z
  x,y,-z
  -y,x-y,-z
  -x+y,-x,-z
  -y,-x,z
  -x+y,y,z
  x,x-y,z
  y,x,z
  x-y,-y,z
  -x,-x+y,z
_cell_length_a                    4.9443
_cell_length_b                    4.9443
_cell_length_c                    500.0000
_cell_angle_alpha                 90.0000
_cell_angle_beta                  90.0000
_cell_angle_gamma                 120.0000
loop_
_atom_site_label
_atom_site_type_symbol
_atom_site_fract_x
_atom_site_fract_y
_atom_site_fract_z
_atom_site_U_iso_or_equiv
_atom_site_adp_type
_atom_site_occupancy
C001    C    -0.33600   -0.16800    0.09050    0.00000  Uiso   1.00
C013    C     0.33333   -0.33333    0.08851    0.00000  Uiso   1.00
NB017   Nb    0.00000    0.00000    0.00000    0.00000  Uiso   1.00
```

## 4.8   BLG-1Ta: BLG-intercalated with 1 Ta

```
data_BLG-1Ta
_symmetry_space_group_name_H-M    'P6/MMM'
_symmetry_Int_Tables_number       191
_symmetry_cell_setting            hexagonal
loop_
_symmetry_equiv_pos_as_xyz
  x,y,z
  -y,x-y,z
  -x+y,-x,z
  -x,-y,z
  y,-x+y,z
  x-y,x,z
  y,x,-z
  x-y,-y,-z
```

```
    -x,-x+y,-z
    -y,-x,-z
    -x+y,y,-z
    x,x-y,-z
    -x,-y,-z
    y,-x+y,-z
    x-y,x,-z
    x,y,-z
    -y,x-y,-z
    -x+y,-x,-z
    -y,-x,z
    -x+y,y,z
    x,x-y,z
    y,x,z
    x-y,-y,z
    -x,-x+y,z
_cell_length_a                  4.9465
_cell_length_b                  4.9465
_cell_length_c                  500.0000
_cell_angle_alpha               90.0000
_cell_angle_beta                90.0000
_cell_angle_gamma               120.0000
loop_
_atom_site_label
_atom_site_type_symbol
_atom_site_fract_x
_atom_site_fract_y
_atom_site_fract_z
_atom_site_U_iso_or_equiv
_atom_site_adp_type
_atom_site_occupancy
C001   C    -0.33620  -0.16810   0.08858   0.00000  Uiso   1.00
C013   C     0.33333  -0.33333   0.08610   0.00000  Uiso   1.00
TA017  Ta    0.00000   0.00000   0.00000   0.00000  Uiso   1.00
```